# The machine learning platform for developers of large systems

Authors[1]: Alexey Naikov[*,**], Anatoly Oreshkin[*,**], Alexey Shvetsov[*,**], Andrey Shevel[*,**]

[*] ITMO University, Saint-Petersburg, Russia

[**] Konstantinov Petersburg Nuclear Physics Institute of NRC "Kurchatov Institute", Gatchina, Russia

{av.naikov,anatoly.oreshkin,alexey,andrey.shevel}@itmo.ru

**Key words:** Retrieval Augmented Generation, Machine Learning, Artificial Neural Networks, Streamlit

**Abstract**

The machine learning system in the form of Retrieval Augmented Generation (RAG) has developed steadily since about 2021. RAG could be observed as a version of the knowledge transfer. In the studied case, the large computing systems are observed as the application point of RAG, which includes large language model (LLM), as a partner for the developing team. Such an approach has advantages during the development process and further in exploitation time.

**Keywords**: artificial neural network, retrieval augmented generation, large language model, computing system development.

**Introduction**

Developing new computing systems is a complex endeavor due to changing technology, constantly evolving requirements during development and system maintenance during their lifetime. The high volume of maintenance work after deployment includes troubleshooting, patching, updating, and modifying components to accommodate new features or security requirements. Investigating unusual events might include scanning system descriptions, the archives of administrator records, administrative orders, official recommendations, system logs, etc. The main aim is to keep the investigation time within reasonable limits.

The progress in artificial neural networks (ANNs) and large language models looked promising approaches to address the above challenges. The appropriate architecture for achieving this is Retrieval-Augmented Generation (RAG) [1,2]. RAG combines the strengths of large language models (LLMs) with specific knowledge about the local system. Such a service is available on the Internet; however, not every development authority might permit to send all technical details to a remote Internet portal. Local RAG is also handy in security contexts, where real-time access to local system logs, administrator experience records, and detailed component descriptions is essential for accurate analysis and decision-making. At the same time, those data must not be desired with any system outside the local organization. RAG includes several components:

---

[1] A. Naikov, A.Oreshkin – valuable efforts to prepare the description, questions and estimate the answers.
A.Shvetsov – valuable efforts to configure the Linux server, A. Shevel – proposal idea, LLM & embedding tools selection, program development.

External Knowledge Source (local documents), Embedding Model which converts the query and local documents into vectors; Retriever, which searches the data in documents most relevant to a user query; Language Model (generator), which generates the final answer taking into account user query, the most relevant part of local documents, and Prompt Template, which instructs the language model (LM) what to do.

In general, it is possible to use a simple schema:

- The administrator can enter a question (or statement) by typing one in natural language.
- The LM generates the answer (inference) in natural language, using data from local databases and archives.

In [3], some experience with RAG architecture in computing networks was observed, and [4] has even more advanced ideas. It seems important to determine the main conditions under which simple (naïve) RAG architecture would be helpful.

**RAG testing and lessons**

In initial tests, it was implemented a naive RAG model utilizing open-source LMs like LLaMA2/3 [5], mistral [6], and others on the platform ollama [7] with limited local data consisting of just a description of computing network segment (CNS). Such the naïve RAG was used together with the CNS description to help the developer or administrator find the required answers to the questions. The CNS descriptions in pdf were placed in a dedicated directory called "docs." The RAG model was started on a Linux server with four simple GPUs, "GeForce GTX 1080 Ti". The several open embedding tools were taken from https://huggingface.co/.

The procedure was general: First, the vector database was built using the results of embedding PDF files in the docs. Then, the administrator's interaction with the RAG architecture was implemented using the front-end package Streamlit [8] and the administrator's web browser. Answers/inferences from the RAG model were returned to the administrator's browser. The administrator could enter the question in natural language, and RAG would attempt to generate a response based on the CNS description. The description was in the form of pdf files, a little bit more than a hundred pages. The initial program snippet was obtained from github.com [9]. The final program text has been significantly redeveloped.

Initially, around ten questions were prepared and entered in the browser one by one in the sequence "enter and send question—wait for the answer." Initially, the answers did not look promising: some were erroneous, and others did not look completely right. Also, several hallucinations occurred in which the model provided incorrect answers without flagging its inability to find the relevant data.

Subsequent analysis of the RAG architecture inferences showed that not all parts of the CNS description were correct, clear, and complete as they should be. For example, when the description authors mistook some details considered "obvious for everybody" and did not include them in the description, the RAG model failed to provide accurate responses. Several parts of the description were corrected and edited to make the content more complete and clearer. The generation temperature was set to the value of 0.1 (the lower the temperature, the fewer hallucinations). In addition, the prompt was edited as well to make instructions for LLM clearer. After that, the previous questions were sent again to RAG architecture, and answers were obtained.

The loop [please see Figure 1] was repeated to generate more adequate answers: enter questions, generate answers, estimate the correctness of the answers, and refine the descriptions until acceptable responses were achieved. Each time, the developers assessed the quality of the answers based on their own opinions regarding whether they were acceptable. If the answer was deemed acceptable, the next question was sent. If the answer was found unacceptable, the developers would begin discussing what could be improved in the description and the prompt. Refining the system description and addressing other gaps significantly improved inference quality. In the testing described, approximately ten loops were required along with the questions. The sufficient number of loops may only be determined through discussion with the developers' team. The usual criterion is to attain correct answers to all prepared questions. The selection of the embedding tool, LLM model, prompt, number of test questions, number of loops, and other parameters also falls under the developers' responsibility.

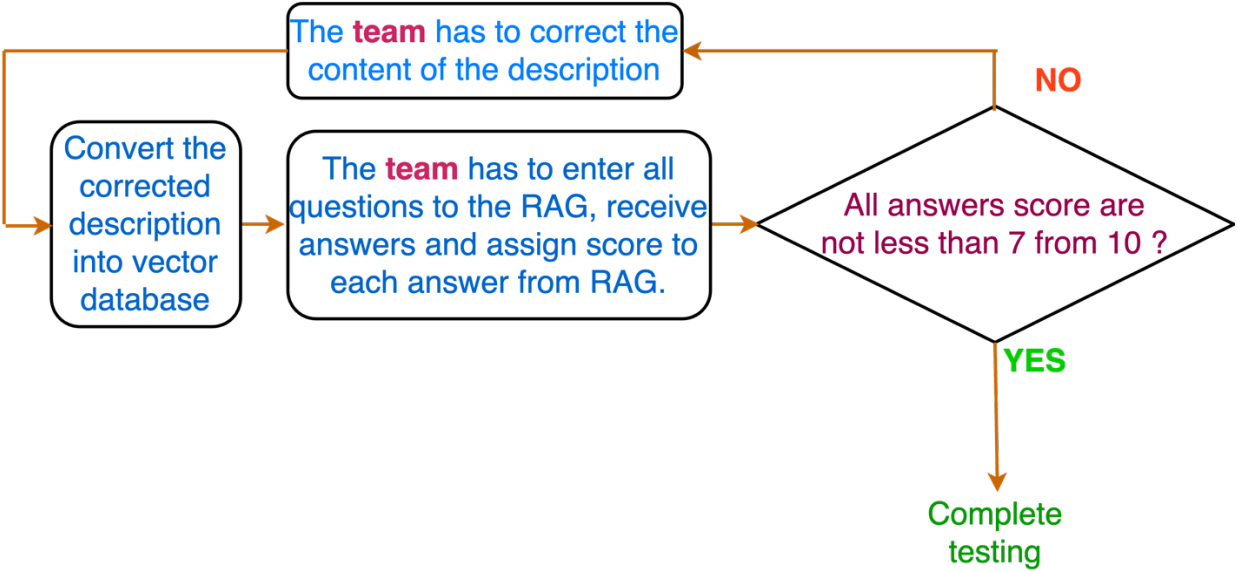

*Figure 1. The testing loop for the system description.*

During the tests, it was observed that the quality of the inferences and answers depended on several factors: the embedding model, the language model, the prompt, and especially on the completeness and clarity of the CNS description. Several screenshots of the output screen are shown at the end of the paper.

To enhance the overall accuracy of the RAG architecture, a loop process of asking test questions, identifying incorrect responses, and refining the description became necessary. Analyzing erroneous or incomplete answers from RAG led to a discussion among the CNS developers' team. The collaborative interactions between the team and RAG proved to be very helpful. Additionally, the question regarding the minimum number of pages in the description of the ready-to-use prototype is intriguing, as it may be beneficial. It was determined that the suggested approach is advantageous for this paper, which consists of 5 pages in the prototype and in

notebooklm.google.com. As the number of pages in documents increases, the expected effect also rises.

In our case, we had a description of the concrete CNS consisting of 150 pages in A4 format. To obtain sufficient confidence in the reliability of the RAG architecture, 50 to 100 test questions should be prepared. It is assumed that if correct answers are provided for these questions, the architecture will be able to accurately answer many other questions not included in the set of test questions. The developers themselves can compose the questions. Obviously, some questions can be generated using any appropriate LLM, such as llama3.x or deepseek-r1. With the mentioned LLMs, one could enter, "*Please prepare the 70 test questions for the description of a server network to estimate the quality of the description.*" Some questions that do not apply to the description could be removed. Test questions should be entered into the RAG architecture, and the answers received should be assessed by the development team using a numerical rating from 0 to 10 (0 signifies an unacceptable answer, while 10 indicates an excellent answer). The correction of the project or system description should continue until the minimum score for the answers to all applicable questions is no less than 7 out of 10. On the current server with old GPUs, some answers were received after almost 10 minutes or so after entering the question, meaning that 50 questions might require several hours. It's much better to get the answer after several seconds rather than minutes.

In the loop was done following:

- The wording has been clarified, and several additional sections have been added to the text description.
- The developers have repeatedly discussed what it is and its origin.
- In other words, the RAG architecture served as a technical aid for developers.

The result is a significantly improved version of the CNS description.

During the improvement of the description, the volume in pages has increased by about 30% (we started with around 100 pages). Generally speaking, anyone could start with the short abstract of the description.

**Integrated Development of computing system and RAG architecture**

The observations from our initial testing suggest a significant opportunity for improvement by developing the new computing system description and the RAG architecture interactively and in parallel. This integrated approach would enable developers to refine the system documentation and development ideas in real time as the RAG architecture is tested, ensuring that the system description and RAG evolve together. Any changes in the system must be entered into the description, which should then be introduced into the RAG. The suggested approach is intriguing for the developer team of large systems. Better results might be expected if the RAG architecture is more advanced than naive RAG, incorporating useful AI agents to analyze the system logs. Analyzing system logs should enhance responses by considering the actual state of the system. Deploying a dedicated server with several GPUs, a library of LLMs, embedding tools, etc., as a RAG installation for the developer team is a worthwhile idea. Options for similar goals are available [10,11]. However, not all local data may be shared outside the developer team.

Additionally, it is evident that specific development will require specific RAG architecture; for example, a local specification may necessitate a particular set of language models.

It is easy to foresee that ready-to-use RAG will soon become a mandatory tool alongside a standard document set for any large system development, providing benefits at multiple stages:

- Initial ideas description and documentation consistency: ensuring that ideas and system descriptions remain consistent, complete, and up to date as the system evolves.
- It's much easier to involve more development team members.
- Maintenance and support when the system is in production: assisting system administrators by providing relevant, real-time information about system status, problems, and recommended actions.
- Future system upgrades: when future upgrades or modifications are necessary, RAG's capability to engage with well-maintained system documentation will simplify these processes.
- There is no need to send local data outside the organization where RAG installation has been deployed.

Several screenshots from different LLMs appear in the section titled "The screenshot examples from RAG output" below. These screenshots display several important parameters for generation.

- RAG_model – the name of the LLM.
- RAG_embed_model – the na e of model for embedding.
- RAG_temperature – the temperature of LLM generation (mainly words and relationships) from text must be used).
- RAG_top_p sets a cumulative probability threshold for selecting the next token during text generation.
- RAG_num_ctx - defines the maximum number of tokens that a language model can "see" or process at once; this is often referred to as the context window or context length.
- RAG_template – text of the prompt. It is important to note that when asking questions and obtaining answers, you can use the natural language that is more convenient for comprehension, rather than the one in which the documents are written.
- RAG_work_dir – working catalog.
- RAG_docs – catalog with documents.
- You entered – the text of entered question.
- Next line shows the name of file with document.
- Very important notification "Please check any answer from RAG!"

For a specific domain, being experienced with all the parameters mentioned might help achieve the required quality of answers. The main condition for success is obviously an advanced team. Even when RAG doesn't provide trivial answers, only the team can estimate the value of such an answer.

**Conclusion**

The breakthrough of this approach lies in the symbiotic development of RAG architecture alongside the system being developed, as specific development will necessitate specific RAG tuning. Creating a direct feedback loop between system development and RAG inference streamlines both processes. Continually refining documentation based on RAG's output helps ensure that the system is well-documented and that the model yields high-quality, relevant information. This reduces the long-term burden of system maintenance, enhances system reliability, and lowers the risk of incorrect or incomplete answers during operation.

In short, the use of RAG architecture in development or administration simplifies/reduces:

- Time to develop a description.
- Attracting new development/administration participants.
- Maintenance of the developed CNS.
- Further modernization during operation.
- Creation of a digital twin of the system being developed (or developed).
- Discussions within the development team (or admins) to clarify the RAG architecture create greater understanding within the team and often generate new ideas.

The approach outlined could benefit teams involved in large-scale technical or scientific development, or an administration team responsible for maintaining substantial physical equipment. Future research in this area will undoubtedly be necessary.

**References**

1. Hao Yu, Aoran Gan, Kai Zhang, Shiwei Tong, Qi Liu and Zhaofeng Liu // Evaluation of Retrieval-Augmented Generation: A Survey // arXiv:2405.07437 [cs.CL] (or arXiv:2405.07437v2 [cs.CL] for this version) https://doi.org/10.48550/arXiv.2405.07437
2. Wenqi Fan et al // A Survey on RAG Meeting LLMs: Towards Retrieval-Augmented Large Language Models // arXiv:2405.06211 [cs.CL] (or arXiv:2405.06211v3 [cs.CL] for this version) https://doi.org/10.48550/arXiv.2405.06211
3. Amar Abane, Abdella Battou, Mheni Merzouki // An Adaptable AI Assistant for Network Management // NOMS 2024-2024 IEEE Network Operations and Management Symposium | 979-8-3503-2793-9/24/$31.00 ©2024 IEEE | DOI: 10.1109/NOMS59830.2024.10574957
4. Ruichen Zhang et al // Interactive AI with Retrieval-Augmented Generation for Next Generation Networking // arXiv:2401.11391 [cs.NI] (or arXiv:2401.11391v1 [cs.NI] for this version) https://doi.org/10.48550/arXiv.2401.11391
5. Hugo Touvron et al // LLaMA: Open and Efficient Foundation Language Models // https://doi.org/10.48550/arXiv.2302.13971.
6. Albert Q. Jiang et al // Mistral 7B // https://doi.org/10.48550/arXiv.2310.06825.
7. Ollama https://ollama.com [last visited 2024-10-28].
8. A faster way to build and share data apps. // https://github.com/streamlit // [last visited 2024-10-28].

9. https://gitlab.com/rahasak-labs/iollama [visited 2025-01-23]
10. GPT that Reviews Your Research Proposal https://peerrecognized.com/gpt-reviews-your-research-proposal/ [visited 2024-12-07]
11. NotebookLM https://notebooklm.google.com [visited 2024-12-07]

**The screenshot examples from RAG output**

```
RAG_model: deepseek-r1:32b
RAG_embed_model: multilingual-e5
RAG_temperature: 0.1
RAG_top_p: 0.5
RAG_num_ctx: 131072
RAG_template: csd_template_eng.txt
RAG_work_dir: /Users/andreyshevel/IOLAAMA-2024-07-25/iollama
RAG_docs: csd
You entered: 2025-05-18 19:14:02 -- what will happen if service SGE is down ?
-- services-notes-2025-05-06.pdf
Completion code: 200

Please check any answer from RAG!

2025-05-18 19:18:02 -- The RAG answer: { "answer": "<think> Okay, I need to figure out what happens if the Service Grid Engine (SGE) goes down. Let me start by recalling what I know about SGE from the provided documents.

First, looking at page 148, there's a script testing server accessibility. It mentions servers and mailing lists for notifications. If SGE is part of these services, its downtime would trigger alerts. But how does that affect operations?

On page 37, it talks about GE-2011.11p1 issues. They mention problems with NIS on virtual machines, which can prevent job submissions. So if SGE is down, users might not be able to submit or run jobs, leading to delays in processing tasks.

Page 43 discusses SGE 8.1.10 setup and commands like qsub for submitting jobs. If the service is down, these commands would fail, preventing any new job submissions. Also, ongoing jobs might be interrupted, causing data loss or incomplete processes.
```

*Figure 2. Screenshot of the output for deepseek (beginning).*

> Page 36 mentions pcfarm-sl6 with GE-2011.11p1 and how users get info via 'news batch'. If SGE is down, this information wouldn't be available, leaving users unaware of the status or unable to manage their jobs.
>
> From page 43 again, if a job fails, qstat shows errors, but without SGE running, even checking statuses becomes impossible. Admins can't use qmon to troubleshoot, so resolving issues would take longer.
>
> Cron jobs on pages 145 and 149 might include scripts that depend on SGE. If SGE is down, these automated tasks could fail, affecting data backups or monitoring processes.
>
> In summary, if SGE goes down, it disrupts job submissions, processing, user access to information, and potentially other dependent services. It's crucial for admins to monitor SGE health and have recovery plans in place. </think>
>
> If the Service Grid Engine (SGE) is down, several critical consequences may occur based on the provided documentation:
>
> 1. **Job Submission Failure**: Users will be unable to submit new jobs using commands like `qsub`. This halts any batch processing tasks that rely on SGE.
>
> 2. **Interrupted Processing**: Any ongoing jobs running through SGE would likely terminate or become unmanageable, potentially leading to data loss or incomplete results.
>
> 3. **Loss of Job Monitoring and Management**: Users won't be able to check job statuses using `qstat` or manage their tasks with `qmon`. This lack of visibility can hinder troubleshooting and resource management.
>
> 4. **Impact on Cluster Utilization**: Since SGE is used for distributing workloads across clusters (as mentioned on page 11), its downtime would render the cluster's computational resources effectively unusable for parallel processing tasks.
>
> 5. **Potential Cron Job Failures**: If any cron jobs depend on SGE operations, they might fail or produce incorrect results, affecting automated processes like backups or monitoring scripts.
>
> 6. **Notification and Alerting**: The system may send out alerts (as seen in the script on page 148) indicating that SGE services are unavailable, prompting administrators to investigate and restore the service.
>
> To mitigate these issues, it's essential for system administrators to monitor SGE health closely and have recovery procedures in place to minimize downtime." }

*Figure 3. Screenshot of the output for deepseek (continuation).*

**RAG_model:** qwen3:30b-a3b

**RAG_embed_model:** multilingual-e5

**RAG_temperature:** 0.1

**RAG_top_p:** 0.5

**RAG_num_ctx:** 131072

**RAG_template:** csd_template_eng.txt

**RAG_work_dir:** /Users/andreyshevel/IOLAAMA-2024-07-25/iollama

**RAG_docs:** csd

**You entered:** 2025-05-18 19:05:38 -- what will happen if service SGE is down ?

-- services-notes-2025-05-06.pdf

Completion code: 200

**Please check any answer from RAG!**

2025-05-18 19:07:38 -- The RAG answer: { "answer": "<think> Okay, the user is asking what happens if the SGE service is down. Let me look through the provided documents to find relevant information.

First, I'll check the pages mentioned. Page 36 talks about SGE-2011.11p1 and mentions that only specific nodes like vrt-04 and vrt-05 are available for batch jobs. If SGE is down, those nodes wouldn't accept tasks. Page 37 mentions using qstat and qmod commands, which would fail if SGE is down. There's also a note about NIS issues on vrt-xx, which could prevent access if SGE is down.

Page 43 discusses SGE 8.1.10 and commands to start/stop it. If SGE is down, users can't submit jobs, and qstat would show errors. The script on page 148 checks server accessibility, which might include SGE servers. If SGE is down, the script would log a problem.

I need to structure the answer with enumerated points, citing the page numbers. Make sure to mention job submission issues, error detection via qstat, possible NIS problems, and the script's role in monitoring. Also, include the manual restart steps from pages 36 and 43.

*Figure 4. Screenshot of the output for qwen3 (beginning).*

> So, the key points are: job submission failures, error detection with qstat, NIS/DNS dependencies, scripted monitoring, manual restart, and impact on specific nodes. Each point should reference the relevant pages.
>
> I think that covers all the necessary information from the provided context. Now, structure it clearly with bullet points and page citations. </think>
>
> If the **SGE (Sun Grid Engine)** service is down, the following outcomes and behaviors are documented in the provided materials:
>
> ## 1. Batch Job Submission and Execution Failures
>
> - **SGE is essential for scheduling and executing batch jobs.** If it is unavailable, users cannot submit or run jobs on the cluster.
> - On **pcfarm-sl6.pnpi.spb.ru** (running GE-2011.11p1), only specific nodes (e.g., `vrt-04`, `vrt-05`) are designated for batch jobs. If SGE is down, these nodes will not process tasks.
> *(Page 36, Page 11)*
>
> ## 2. Error Detection via `qstat` and `qmon`
>
> - The `qstat` command (used to check job status) will fail or return errors if SGE is down.
> - Administrators can use `qmon` (a GUI tool) to monitor SGE, but it will not connect to the scheduler if the service is inactive.
> *(Page 37, Page 43)*

*Figure 5. Screenshot of the output for qwen3 (continuation-1).*

## 3. Dependency on NIS and DNS

- SGE relies on **NIS (YP)** and **DNS** for user authentication and node resolution. If SGE is down, issues with NIS (e.g., `ypserv` failures on `pcfarm-new`) could prevent access to compute nodes like `vrt-xx`.
- For example, if `ypserv` is misconfigured or down, SGE jobs may fail to authenticate users or resolve node names.

(Page 36, Page 11)

## 4. Scripted Monitoring and Alerts

- The script `/usr/local/bin/TestDiskUsage.bash` (referenced on **Page 148** and **Page 149**) checks system health but does not directly monitor SGE. However, if SGE is down, it could indirectly trigger disk usage alerts if jobs fail to complete.

Figure 6. Screenshot of the output for qwen3 (continuation-2).

- Cron jobs (e.g., `4 9,14,18 * * * /usr/local/bin/TestDiskUsage.bash`) might log errors if SGE-related processes are unresponsive.

## 5. Manual Restart Requirements

- To restore SGE, administrators must manually restart the service:
  - For **GE-2011.11p1**:

    ```
    /etc/init.d/sgemaster start
    ```

  - For **SGE 8.1.10**:

    ```
    /etc/init.d/sgemaster start
    ```

(Page 36, Page 43)

Figure 7. Screenshot of the output for qwen3 (continuation-3).

## 6. Impact on Specific Nodes

- On **pcfarm-sl6.pnpi.spb.ru**, SGE is tied to specific virtual machines (e.g., `vrt-01-sl6.7-x86_64`). If SGE is down, these nodes will not process jobs, even if the underlying hardware is functional. *(Page 36)*

## Summary

If SGE is down, **batch job execution halts, error detection tools like `qstat` fail, and dependencies on NIS/DNS may exacerbate access issues**. Administrators must restart SGE services and verify node connectivity to restore functionality.

**Relevant Pages**: 11, 36, 37, 43, 148, 149." }

*Figure 8. Screenshot of of the output for qwen3 (continuation-4).*